\documentclass[10pt]{article}
\usepackage[dvips]{graphicx}

\usepackage{epsfig}
\usepackage{graphicx}
\usepackage{indentfirst}
\usepackage{amsmath}
\usepackage{amsfonts}
\usepackage{amssymb}

\begin{document}
\begin{center}
{\Large\bf   Thermodynamics in Little Rip cosmology in the framework of a type of $f(R, T)$ gravity \\}
\medskip

M. J. S. Houndjo$^{a,b}$\footnote{e-mail:
sthoundjo@yahoo.fr}, \, F. G. Alvarenga$^{a}$\footnote{e-mail:f.g.alvarenga@gmail.com},\, Manuel E. Rodrigues$^{c}$\footnote{e-mail: esialg@gmail.com}, \, Deborah F. Jardim$^{d}$\footnote{dfjardim@gmail.com} \\and R. Myrzakulov$^{(e)}$\footnote{e-mail: rmyrzakulov@csufresno.edu}

$^{a}$ \,{Departamento de Engenharia e Ci\^{e}ncias Naturais - CEUNES -
Universidade Federal do Esp\'irito Santo\\
CEP 29933-415 - S\~ao Mateus/ ES, Brazil}\\
$^b$ \,{\it Institut de Math\'{e}matiques et de Sciences Physiques (IMSP)}\\
 {\it 01 BP 613 Porto-Novo, B\'{e}nin}\\
$^{c}$\,{Universidade Federal do Esp\'irito Santo - 
Centro de Ci\^{e}ncias Exatas - Departamento de F\'isica\\
Av. Fernando Ferrari s/n - Campus de Goiabeiras\\
CEP29075-910 - Vit\'oria/ES, Brazil}\\
$^{d}$\, \ Universidade Federal dos Vales do Jequitinhonha e Mucuri, ICTM\\
Rua do Cruzeiro, 01, Jardim S\~{a}o Paulo\\
CEP39803-371 - Teofilo Otoni, MG - Brazil\\
$^{e}$\, {\ Eurasian International Center for Theoretical Physics\\
L.N. Gumilyov Eurasian National University, Astana 010008, Kazakhstan}

\date{}

\end{center}
\begin{abstract}
Cosmological reconstruction of Little Rip model in $f(R, T)$ gravity is investigated, where $R$ is the curvature scalar and $T$ the trace of the energy momentum tensor. The model perfectly reproduces the present stage of the universe, characterized by the $\Lambda CDM$ model, without singularity at future finite-time (without the Big Rip). The input parameters are determined according to Supernovae Cosmology data and perfectly fit with the WMAP around the Little Rip. Moreover, the thermodynamics is considered in this Little Rip cosmology and it is illustrated that the second law of thermodynamics is always satisfied around the Little Rip universe for the temperature inside the horizon being the same as that of the apparent horizon. Moreover, we show the existence of a stable fixed point in the Little Rip universe which confirms that this is actually a late-time attractor in the phantom-dominated universe. The linear perturbation  analysis is performed around the critical points, showing that the Little Rip model obtained is stable.
\end{abstract}

Pacs numbers: 04.50.Kd, 95.36.+x, 98.80.-k
\section{Introduction}
Recent astronomical data from Type Ia supernovae \cite{1leonardo} as well as from the CMB  spectrum \cite{2leonardo} confirm that our universe is undergoing an accelerated expansion period. In order to comply with this  feature, dark energy content for the universe may be postulated \cite{3leonardo}, with undesired properties, such as the violation of some energy conditions. Other way is to consider extended theories of gravity, which besides their ability to describe the inflation in the early universe \cite{1mohseni}, can be used as candidates to explain the present acceleration of the universe \cite{2mohseni} without the requirement to introduce the exotic matter with negative pressure, dubbed as dark energy. A method for generalizing the Einstein theory is to consider the gravitational Lagrangian as a function of the curvature $R$ and the trace $T$ of the energy momentum tensor, the so-called $f(R, T)$ theory \cite{fRTpaper}. This theory of modified gravity has attracted more attentions and some interesting results have been found \cite{stephaneseul3}-\cite{flavio}.\par
Recently, a novel scenario  has been proposed in \cite{brevik1}, the so-called ``Little Rip"  (LR). Models without a future singularity in which the energy density of the dark energy, $\rho_{DE}$,  increases with time will nonetheless eventually lead to a dissolution of bound structures at some point in the future. In the LR, the scale factor and the density are never infinite at a finite time. Typically, the LR interpolates between the Big Rip (BR), where the scale factor and energy density diverge, and the $\Lambda$CDM model, where there is no such divergence and no disintegration because the dark energy 
remains constant. Mathematically, the LR can be represented as an infinite limite sequence which has the BR and the $\Lambda$CDM as its boundaries. Several papers in literature deal with LR cosmology in GR and the modified $f(R)$ theory of gravity \cite{diegoproposal,odintsovproposal}. \par
In this paper, we consider the process of occurrence of the LR  and propose to investigate the corresponding explicit form  of type $R+2g(T)$ model. Here, we are considering $f(R, T)$ as $f(R, T)=f_1(R)+f_2(T)$, where we specially consider $f_1(R)=R$ and $f_2(T)=2g(T)$.
\par
 
Note that black hole thermodynamics indicates the fundamental connection between the gravitation and thermodynamics \cite{23thermo4,24thermo4}. The Einstein equation was derived from the Clausius relation in thermodynamics with the proportionality of the entropy to the horizon area in General Relativity (GR) \cite{26thermo4}. This procedure was developed to more general extended gravitational theories \cite{27thermo4,28thermo4}, and also in so-called $f(\mathcal{T})$ gravity \cite{bamba,daouda2}, where $\mathcal{T}$ is the torsion scalar. Note also that a new method is now developed for studying the thermodynamics of equilibrium system, the so-called geometrodynamics, and some interesting results have been found \cite{zui}-\cite{deborah2}.\par
In this paper the proposal is the same but in the framework of $R+2g(T)$ gravity, analysing the validity of the law of thermodynamics in a LR model. In this optic, we investigate the non-equilibrium description of thermodynamics near the LR. We show that the second  law of thermodynamics is always verified in LR model  for the temperature inside the horizon being the same as that of the apparent horizon. Moreover, we investigate the linear perturbations around the critical and see that the LR model found in this work is stable\par
The paper is organized as follows. In Sec. $2$, we present the general formalism of $f(R,T)$ gravity; the special $R+2g(T)$ gravity is adopted and the model leading to LR is obtained. In Sec. $3$, the thermodynamics in LR cosmology is considered, where the first and second law are investigated. The stability analysis of $R+2g(T)$ LR model is performed in the Sec. $4$, and the conclusion is presented in the Sec. $5$.

\section{General formalism of f(R, T) gravity}
Let us consider the general model of $f(R, T)$  gravity whose action can be described by the action       
\begin{eqnarray}\label{manuel1}
S=\int d^4x\sqrt{-g} \left\lbrace \frac{1}{2\kappa}f(R, T)+\mathcal{L}_{m}\right\rbrace \,\,,
\end{eqnarray}
where $\kappa=8\pi G$, with $G$ the Newtonian gravitational constant, $R$ is the curvature scalar and $T$ the trace of the energy momentum tensor, defined from the matter Lagrangian density $\mathcal{L}_m$ by
\begin{eqnarray}\label{manuel2}
T_{\mu\nu}=-\frac{2}{\sqrt{-g}}\frac{\delta\left( \sqrt{-g}\mathcal{L}_m\right) }{\delta g^{\mu\nu}}\,\,\,.
\end{eqnarray}
Varying the action $S$ with respect to the metric $g^{\mu\nu}$, one obtains \cite{fRTpaper},
\begin{eqnarray}\label{manuel3}
f_R(R, T)R_{\mu\nu}-\frac{1}{2}f(R, T)g_{\mu\nu}+\left( g_{\mu\nu}\Box -\nabla_\mu\nabla_\nu\right) f_R(R, T)= \kappa T_{\mu\nu}-f_T(R, T)T_{\mu\nu}-f_T(R, T)\Theta_{\mu\nu}\,\,,
\end{eqnarray}
where $\Theta_{\mu\nu}$ is defined by 
\begin{eqnarray}\label{manuel4}
\Theta_{\mu\nu}\equiv g^{\alpha\beta}\frac{\delta T_{\alpha\beta}}{\delta g^{\mu\nu}}= -2T_{\mu\nu}+g_{\mu\nu}\mathcal{L}_m-2g^{\alpha\beta}\frac{\partial^{2}\mathcal{L}_m}{\partial g^{\mu\nu}\partial g^{\alpha\beta}}\,\,.
\end{eqnarray}
Here $f_R$ and $f_T$ denote the derivatives of  $f$ with respect to $R$ and $T$, respectively. 
Let us assume for simplicity that the function $f$ is given by $f(R, T)=f_1(R)+f_2(T)$, where $f_1(R)$ and $f_2(T)$ are arbitrary functions of $R$ and $T$, respectively. Then,  (\ref{manuel3})  can be written as 
\begin{eqnarray}\label{manuel5}
f_{1R}(R)R_{\mu\nu}-\frac{1}{2}f_1(R)g_{\mu\nu}+\left( g_{\mu\nu}\Box-\nabla_\mu\nabla_\nu\right) f_{1R}(R)=\kappa T_{\mu\nu}-f_{2T}(T)T_{\mu\nu}-f_{2T}(T)\Theta_{\mu\nu}+\frac{1}{2}f_2(T)g_{\mu\nu}\,\,.
\end{eqnarray}
Assuming that the matter content is a perfect fluid, the stress tensor is given by
\begin{eqnarray}\label{manuel6}
T_{\mu\nu}=(\rho+p)u_\mu u_\nu-pg_{\mu\nu}\,\,,
\end{eqnarray}
where $u_{\mu}$ is the four-velocity satisfying $u_\mu u^{\mu}=1$. Then,  the matter Lagrangian density can be taken as $\mathcal{L}_m=-p$, and  $\Theta_{\mu\nu}=-2T_{\mu\nu}-pg_{\mu\nu}$. Hence,  Eq. (\ref{manuel5}) becomes
\begin{eqnarray}\label{manuel7}
f_{1R}(R)R_{\mu\nu}-\frac{1}{2}f_1(R)g_{\mu\nu}+\left( g_{\mu\nu}\Box-\nabla_\mu\nabla_\nu\right) f_{1R}(R)=\kappa T_{\mu\nu}+f_{2T}(T)T_{\mu\nu}+\left[ f_{2T}(T)p+\frac{1}{2}f_2(T)\right] g_{\mu\nu}\,\,.
\end{eqnarray}
Now, we propose to assume a special case where $f_1(R)=R$ and $f_2(T)= 2g(T)$, i.e., $f(R, T)= R+2g(T)$ as $2g(T)$ additive term to Einstein-Hilbert one. This case seems interesting and has been widely studied for other purposes \cite{fRTpaper}-\cite{juliano} and interesting results have been obtained. Then, Eq. (\ref{manuel7}) becomes
\begin{eqnarray}\label{manuel8}
R_{\mu\nu}-\frac{1}{2}Rg_{\mu\nu}=\tilde{\kappa} T_{\mu\nu}+\left[2g_T(T)p+g(T)\right]g_{\mu\nu}\,\,\,,\quad\quad \tilde{\kappa}=\kappa+2g_T(T)\,\,\,.
\end{eqnarray} 
We see from this equation that coupling constant $\kappa$ in GR,  becomes a running constant $\tilde{\kappa}$ in $R+2g(T)$ gravity.\par 
Let us now consider the spatially flat  Friedmann-Robertson-Walker (FRW) line element
\begin{equation}\label{manuel9}
ds^2= dt^2-a^2(t)d{\bf x}^2 \,\,,
\end{equation}
where $a(t)$ is the scale factor. One easily gets the first and second generalized Friedmann equations, respectively as 
\begin{eqnarray}
3H^2 &=& \tilde{\kappa}\left[\rho+\frac{1}{\tilde{\kappa}}\left(2pg_T+g\right)\right]\,\,\,,\label{manuel10}\\
-2\dot{H}-3H^2&=&\tilde{\kappa}\left[p-\frac{1}{\tilde{\kappa}}\left(2pg_T+g\right)\right] \,\,\,, \label{manuel11}
\end{eqnarray}
where $H=\dot{a}/a$ is the Hubble parameter and the dote denoting the derivative with respect to the cosmic time $t$. Calling the expressions in brackets in the right side of (\ref{manuel10}) and (\ref{manuel11}) respectively the effective energy density $\rho_{eff}$ and pressure $p_{eff}$, one write (\ref{manuel10}) and (\ref{manuel11}) as
\begin{eqnarray}
3H^2&=&\tilde{\kappa} \rho_{eff}\,\,\,,\label{manuel12}\\
-2\dot{H}-3H^2&=&\tilde{\kappa} p_{eff}\,\,\,, \label{manuel13}
\end{eqnarray}
Note that by combining (\ref{manuel12}) and (\ref{manuel13}), one can write $p_{eff}=-\rho_{eff}-2\dot{H}/\tilde{\kappa}$. Since, the effective energy density is directly linked with the square of the Hubble through (\ref{manuel12}), one can take $\dot{H}$ as function of $\rho_{eff}$. Then, one can write the equation of state as 
\begin{eqnarray}
p_{eff}=-\rho_{eff}-f(\rho_{eff}) \,\,\,\,.\label{manuel14}
\end{eqnarray}
From the conservation law, the effective energy density evolves as 
\begin{eqnarray}
\frac{d(\tilde{\kappa}\rho_{eff})}{dt}&=&-3H\tilde{\kappa}\left(\rho_{eff}+p_{eff}\right)\nonumber\\
&=& 3H\tilde{\kappa}f(\rho_{eff}) \,\,\,.\label{manuel15}
\end{eqnarray} 
Let us now assume that the scale factor can be written as \cite{paul}
\begin{equation}
a(t)=e^{h(t)}\,\,\,,\label{manuel16}
\end{equation}
where $h(t)$ is a non-singular function. Hence, the effective energy density can be written as $\rho_{eff}=3\dot{h}^2/\tilde{\kappa}$. The condition for $\tilde{\kappa}\rho_{eff}$ being an increasing function of the scale factor $a$ is $d(\tilde{\kappa}\rho_{eff})/da=(6/\dot{a})\dot{h}\ddot{h}>0$. This holds as long as $\ddot{h}>0$. By using Eq. (\ref{manuel15}), it appears that  an increasing $\tilde{\kappa}\rho_{eff}$  is ensured by $\tilde{\kappa}f(\rho_{eff})>0$. Integrating Eq. (\ref{manuel15}), one gets
\begin{eqnarray}
a=a_0\exp{\left(\int_{\rho_0}^{\rho_{eff}} \frac{d(\tilde{\kappa}\rho)}{3\tilde{\kappa}f(\rho)}\right)}\,\,\,.\label{manuel17}
\end{eqnarray}
Also, Eq. (\ref{manuel12}) yields
\begin{eqnarray}
t= \int^{\rho_{eff}}_{\rho_0} \frac{d(\tilde{\kappa}\rho)}{\sqrt{3\tilde{\kappa}\rho}\tilde{\kappa}f(\rho)}\,\,\,.\label{manuel18}
\end{eqnarray}
The condition for a big rip singularity is that the integral (\ref{manuel18}) converges. Choosing $f$ such that  $f(\rho)=A(\rho/\tilde{\kappa})^{1/2}$ \cite{paul,brevik}, one gets
\begin{eqnarray}
t=\frac{1}{A\sqrt{3}}\ln{\left(\frac{\tilde{\kappa}\rho_{eff}}{\kappa\rho_0}\right)}\,\,\,. \label{manuel19}
\end{eqnarray}
A simple analysis of (\ref{manuel19}) shows that $\kappa\rho_{eff}$ does not diverge until an infinite time $t$ has elapsed. This is the LR phenomenon. From (\ref{manuel17}), we can express $\rho_{eff}$ in functions of $a$, 
\begin{eqnarray}
\rho_{eff}=\frac{\kappa}{\tilde{\kappa}}\rho_0\left[1+\frac{3A}{2\sqrt{\kappa\rho_0}}\ln{\left(\frac{a}{a_0}\right)}\right]^2\,\,\,,\label{manuel20}
\end{eqnarray}
and also using Eq. (\ref{manuel12}), one obtains after integration 
\begin{eqnarray}
a(t)=a_0\exp{\left\lbrace \frac{2\sqrt{\kappa\rho_0}}{3A}\left[\exp{\left(\frac{A\sqrt{3}}{2}\,\,\,t\right)}-1\right]\right\rbrace}\,\,\,\,,\label{manuel21}
\end{eqnarray}
that we put in a compact form as \cite{makarenko}
\begin{equation}\label{manuel22}
a(t)=a_0\exp{\alpha\left(e^{\beta t}-1\right)}\,\,\,,
\end{equation}
where $\alpha$ and $\beta$ can be easily found  by identifying (\ref{manuel22}) with (\ref{manuel21}), as $\alpha=2\sqrt{\kappa\rho_0}/(3A)$ and $\beta=A\sqrt{3}/2$.
\subsection{Little Rip $R+2g(T)$ type model}

Here, we propose to construct model of type $R+2g(T)$ according to LR scale factor (\ref{manuel22}).\par
With (\ref{manuel22}), the Hubble parameter and its first derivative read
\begin{eqnarray}
H(t)=\alpha\beta e^{\beta t}\,\,\,,\quad \dot{H}=\alpha\beta^2e^{\beta t}\,\,\,\,.\label{manuel23}
\end{eqnarray}
Since our task here is to determine the algebraic action function $g$ in terms of $T$, we have to express in a first time the cosmic time $t$ in function of the trace $T$. The unique way to realize this is considering the equations of continuity related to the ordinary content of the universe, from which the ordinary energy density should be found, and throughout the relation $T=\rho-3p$, $t$ could  be found in function of the $T$. By considering that the interaction between the dark energy and the ordinary content of the universe is $Q$, the equations of continuity for dark energy and ordinary matter can  be written respectively as
\begin{eqnarray}
\dot{\rho}_d+3H\left(\rho_d+p_d\right)=Q\,\,,\label{deborah24}\\
\dot{\rho}+3H\left(\rho+p\right)=-Q\,\,\,, \label{deborah25}
\end{eqnarray} 
where the subscript $``d"$ indicates the dark energy. 
In this work we consider that the interaction is proportional to the the product of the Hubble parameter by the energy density of the ordinary matter, assuming it as $Q=3qH\rho$, where $q$ is a constant. Thus, the equation (\ref{deborah25}) yields
$\rho=\rho_0a^{-3(1+\omega+q)}$, and from (\ref{manuel22}), one gets 
\begin{eqnarray}
T=\rho-3p=\rho_0(1-3\omega)\exp{\left[-3\alpha(1+\omega+q)(e^{\beta t}-1)\right]}\,\,\,,\label{manuel24}
\end{eqnarray}
where the equation of state $p=\omega \rho$ is assumed for the ordinary content of the universe. Substituting the second term of (\ref{manuel23}) into (\ref{manuel24}), one acquires
\begin{eqnarray}
\dot{H}=\alpha\beta^2+\ln\left\{\left[\rho_0(1-3\omega)\right]^{\frac{\beta^2}{3(1+\omega+q)}}T^{\frac{-\beta^2}{3(1+\omega+q)}}\right\}\,\,\,\,.\label{manuel25}
\end{eqnarray}
On the other hand, by summing (\ref{manuel10}) with (\ref{manuel11}), one gets
\begin{eqnarray}
\rho\left(1+\omega\right)\left(\kappa+2g_T\right)+2\dot{H}=0\,\,\,\,\,.\label{manuel26}
\end{eqnarray}
By using (\ref{manuel25})  and the first equality of (\ref{manuel24}), Eq.~(\ref{manuel26}) becomes
\begin{eqnarray}
\frac{1+\omega}{1-3\omega}\left(\kappa+2g_T\right)T+2\alpha\beta^2+2\ln\left\{\left[\rho_0(1-3\omega)\right]^{\frac{\beta^2}{3(1+\omega+q)}}T^{\frac{-\beta^2}{3(1+\omega+q)}}\right\}=0\,\,\,,\label{manuel27}
\end{eqnarray}
whose general solution reads
\begin{eqnarray}
g(T)=-\frac{\kappa T}{2}-\frac{\alpha\beta^2(1-3\omega)}{1+\omega}\ln\left\{\left[\rho_0(1-3\omega)\right]^{\frac{\beta^2}{3(1+\omega+q)}}T^{\frac{-\beta^2}{3(1+\omega+q)}}\right\}\nonumber\\+\frac{3(1-3\omega)(1+\omega+q)}{2\beta^2(1+\omega)}\ln^2\left\{\left[\rho_0(1-3\omega)\right]^{\frac{\beta^2}{3(1+\omega+q)}}T^{\frac{-\beta^2}{3(1+\omega+q)}}\right\}+C_1\,\,\,,\label{manuel28}
\end{eqnarray}
where $C_1$ is an integration constant. The corresponding $R+2g(T)$ model is
\begin{eqnarray}
f(R,T)=R-\kappa T-\frac{2\alpha\beta^2(1-3\omega)}{1+\omega}\ln\left\{\left[\rho_0(1-3\omega)\right]^{\frac{\beta^2}{3(1+\omega+q)}}T^{\frac{-\beta^2}{3(1+\omega+q)}}\right\}\nonumber\\+\frac{3(1-3\omega)(1+\omega+q)}{\beta^2(1+\omega)}\ln^2\left\{\left[\rho_0(1-3\omega)\right]^{\frac{\beta^2}{3(1+\omega+q)}}T^{\frac{-\beta^2}{3(1+\omega+q)}}\right\}+2C_1\,\,\,. \label{manuel29}
\end{eqnarray}
The constant $C_1$ can be determined in the following way. For GR without cosmological constant, one has $3H^2_0=\kappa\rho_0$, where $H_0$ and $\rho_0$ are respectively the current values of the Hubble parameter and energy density. From the observational data, the current value of the Hubble parameter is $H_0=2.1\times 0.7\times 10^{-42}$GeV \cite{51thermo2,secondde3thermo2}. In this general case, working with $f(R, T)$ model, the initial condition has to be the same, that is $3H_0^2=\tilde{\kappa}\rho_{eff}|_{t=t_0}$. This leads to
\begin{eqnarray}
2\rho_0(1+\omega)g_T(T_0)+g(T_0)=0\,\,\,\,,\label{manuel30}
\end{eqnarray}
where $T_0=\rho_0(1-3\omega)$ is the current trace of the energy momentum tensor. By 
using (\ref{manuel30}) and (\ref{manuel28}), one gets
\begin{eqnarray}
C_1=\frac{\kappa(3-\omega)T_0}{2(1-3\omega)}-\frac{2\alpha\beta^4}{3(1+\omega+q)}
\,\,\,.\label{manuel31}
\end{eqnarray}
Furthermore, by using the initial condition $\kappa\rho_0=3H^2_0$, one obtains $\alpha=2H_0/(A\sqrt{3})$ and $\beta=A\sqrt{3}/2$, which leads to $\alpha\beta=H_0$. In \cite{paul}, the parameter $A$, which also appears here is Eq.~(\ref{manuel19}) is chosen to make a best fit to the lasted supernova data \cite{20paul}, and can be found for the range bounded as $2.74\times 10^{-3}$ Gyr$^{-1}\leq A\leq 9.67\times 10^{-3}$Gyr$^{-1}$. We also have the constant  $\kappa=8\pi G = 8\pi /M_{Pl}$, with the Plank mass of $M_{Pl}=G^{\,-1/2} = 1.2\times 10^{19}$GeV.  Then, the parameters $\alpha$ and $\beta$ can perfectly be found according to observational data. \par 

By calculating the parameter of effective equation of state, one obtains
\begin{eqnarray}
\omega_{eff}&=&\frac{\tilde{\kappa}p+ 2pg_T(T)+g(T)}{\tilde{\kappa}\rho+ 2pg_T(T)+g(T)}\nonumber\\
&=&-1+\frac{(\rho+p)\tilde{\kappa}}{\tilde{\kappa}\rho+ 2pg_T(T)+g(T)}\nonumber\\
&=&-1-\frac{2}{3\alpha}e^{-\beta\,t}\label{manuel32}\,\,.
\end{eqnarray}
It is easy to observe from (\ref{manuel32}) that for any $t$ from this present time $t_0$ to the future, $\omega_{eff}<-1$. This means that the universe is always in phantom phase. One can also observe that in the limit of large $t$, ($t\rightarrow \infty$), $\omega_{eff} \rightarrow -1$ and hence the LR scenario can hold.\par
An interesting feature to be noted in this model is that, from (\ref{manuel22}), for small value of the cosmic time, meaning that we are around the present stage  of the universe\footnote{The small values of time are values closed to the present time $t_0$, since, $t$ may be substituted by $t-t_0$ such that for $t$ closed to $t_0$, the difference $t-t_0$ is small.}, one gets $e^{\beta t}-1\sim \beta t$ and then, $a(t)\propto e^{\alpha(e^{\beta t}-1)}\sim e^{\alpha\beta t} =e^{H_0t}$. In this situation, $\dot{H}=0$ and $\omega_{eff}=-1$; this is the  $\Lambda CDM$ model,  where the gravitational action reads $R-2\Lambda$, with $\Lambda$ the cosmological constant. \par 
In general this $R+2g(T)$ may be view as a type of the so-called $\Lambda(t)$ gravity \cite{nikodem}, where $\Lambda(t)=2g[T(t)]$, and this is an interesting model because it provides the interaction between matter and dark energy \cite{4nikodem}. This model has an advantage over $f(R)$ gravity theories because we do not need to transform (conformally) from a metric to another, and there is no question about which metric is physical since we use one and the same metric tensor \cite{10nikodem}.
\par
Without loss of generality, let us assume that the Hubble parameter can be written as \begin{eqnarray}\label{manuel33}
H(t)=H_{LR}e^{-\beta(t_{LR}-t)}\,\,\,,
\end{eqnarray}
from which the scale factor can be written as
\begin{eqnarray}\label{manuel34}
a(t)=a_{LR}\exp{\left[\frac{H_{LR}}{\beta}\left(e^{-\beta(t_{LR}-t)}-1 \right)\right]}\,\,\,,
\end{eqnarray}
where $H_{LR}$ and $a_{LR}$ denote respectively  the Hubble  parameter and the scale factor at LR time $t_{LR}$. Note that in the LR model, even if $H$ is finite in finite future, in the same, when $H$ becomes infinite, the inertial force becomes very strong and destroys  any bounded object \cite{brevik}. This has been dubbed as LR \cite{paul}. For the model of type (\ref{manuel33}), the time $t_{LR}-t_0$ from the present until the destruction of the Earth-Sun system has been estimated to $t_{LR}-t_0 \sim 146 $Gyrs \cite{brevik}.\par

Then, one has the model that describes very well the present accelerating universe, and presents any finite time singularity in future, i.e., the Big Rip.


\section{Thermodynamics in LR cosmology}
Here, we present thermodynamics in $f(R, T)$ and try to check the viability of such a theory. By following the same procedure in Refs. \cite{23bamba, 55bamba}, we examine whether the second law of thermodynamics can be verified near the LR. In GR, the proportionality of the entropy to the horizon area allows to derive the Einstein equation from the Clausius relation in thermodynamics \cite{58bamba}. Consequence of this has been applied to more general extended gravitational theories \cite{59bamba,60bamba}, and we propose to perform this in $f(R, T)$ gravity around the LR.
\subsection{First Law of Thermodynamics}
According to the recent type Ia Supernovae data, it is suggested that in the accelerating universe the enveloping surface should be the apparent horizon rather than the event one from the thermodynamic point of view \cite{40thermo3}.
In flat FRW spacetime, the radius at the apparent horizon is given by $\tilde{r}_A=1/H$ \cite{bamba}, and the dynamical apparent horizon is determined by the relation
\begin{eqnarray}
h^{\alpha\beta}\partial_{\alpha}\tilde{r}\partial_{\beta}\tilde{r}=0\,\,,\quad h_{\alpha\beta}=diag\left(1, -a^2(t)\right)\,\,\,.\label{manuel38}
\end{eqnarray}
The first derivative of the $\tilde{r}_A$, leads to 
\begin{eqnarray}
d\tilde{r}_A=-\tilde{r}^3_A\dot{H}Hdt\,\,\,.\label{manuel39}
\end{eqnarray}
On the other hand, the equations (\ref{manuel12}) and (\ref{manuel13}) can be put into the following set of equations
\begin{eqnarray}
H^2&=&\frac{\tilde{\kappa}}{3}\rho_{eff}\,\,\,,\label{manuel40}\\
\dot{H}&=&-\frac{\tilde{\kappa}}{2}\left(\rho_{eff}+p_{eff}\right)\,\,.\label{manuel41}
\end{eqnarray}
By making use of  (\ref{manuel41}), one can write (\ref{manuel39}) as 
\begin{eqnarray}
d\tilde{r}_A=\frac{\tilde{\kappa}}{2}\tilde{r}_A^3H\left(\rho_{eff}+p_{eff}\right)dt\,\,\,.\label{manuel42}
\end{eqnarray}
In GR, the Berkenstein-Hawking horizon entropy is expressed as $S_{BH}=\mathcal{A}/(4G)$ \cite{35thermo1,56bamba}, where $\mathcal{A}=4\pi \tilde{r}^2_A$ is the area of the apparent horizon. Note from Eq.~(\ref{manuel8}) that the Newtonian gravitational constant $G$ is replaced by the running gravitational coupling parameter $\tilde{G}=G+2g_T(T)/(8\pi)$. Then, in the context of this special modified gravity, as in \cite{35thermo1}, the entropy is defined as Wald entropy and reads $S=\mathcal{A}/(4\tilde{G})$.
By using (\ref{manuel42}), the derivative of the horizon entropy is written as
\begin{eqnarray}
dS=8\pi^2r^4_A H\left(\rho_{eff}+p_{eff}\right)dt-\frac{\pi \tilde{r}^2_A}{\tilde{G}^2}d\tilde{G}\,\,\,.\label{manuel43}
\end{eqnarray}
One can now calculate the Hawking temperature $T_H=|\sigma_{sg}|/(2\pi)$, corresponding to the associated temperature of the apparent horizon, where $\sigma_{sg}$ denotes the surface gravity which is given by \cite{62bamba}
\begin{eqnarray}
\sigma_{sg}&=&\frac{1}{2\sqrt{-h}}\partial_{\alpha}\left(\sqrt{-h}h^{\alpha\beta}\partial_{\beta}\tilde{r}\right)\,\,\,\nonumber\\
&=&\frac{1}{\tilde{r}_A}\left(\frac{\dot{\tilde{r}}_A}{2H\tilde{r}_A}-1\right)\nonumber\\
&=&-\frac{\tilde{r}_A}{2}\left(\dot{H}+2H^2\right)\nonumber\\
&=&-\frac{\tilde{\kappa}\tilde{r}_A}{12}\left(1-3\omega_{eff}\right)\rho_{eff}\,\,\,,\label{manuel44}
\end{eqnarray}
where we used $p_{eff}/\rho_{eff}=\omega_{eff}$. We clearly see from (\ref{manuel44}) that the surface gravity is positive for $\omega_{eff}>1/3 $. Hence, the temperature is obtained as 
\begin{eqnarray}
T_H=\frac{1}{2\pi\tilde{r}_A}\left(1-\frac{\dot{\tilde{r}}_A}{2H\tilde{r}_A}\right)\,\,\,.\label{manuel45}
\end{eqnarray}
Now, by combining (\ref{manuel45}) and (\ref{manuel43}), we get
\begin{eqnarray}
T_HdS= 4\pi\tilde{r}^{3}_{A}H\left(\rho_{eff}+p_{eff}\right)dt-2\pi \tilde{r}^{3}_{A}H\left(
\rho_{eff}+p_{eff}\right)d\tilde{r}_{A}+\frac{\tilde{r}_{A}\dot{\tilde{G}}}{4\tilde{G}^2}d\tilde{r}_A-\frac{\tilde{r}_A}{2\tilde{G}^2}d\tilde{G}\,\,\,. \label{manuel46}
\end{eqnarray}
Defining the Misner-Sharp energy as 
\begin{eqnarray}
E=\frac{\tilde{r}_A}{2\tilde{G}}= V \rho_{eff}\,\,\,, \label{manuel47}
\end{eqnarray}
where $V=4\pi \tilde{r}^3_A/3$. 

 From the equation of continuity, one acquires
$d(\rho_{eff})/dt+3H\left(\rho_{eff}+p_{eff}\right)=
-3H^2\dot{\tilde{\kappa}}/(\tilde{\kappa}^2)$. By using this, one gets
\begin{eqnarray}
dE=4\pi\tilde{r}^{3}_{A}H\rho_{eff}d\tilde{r}_A-4\pi\tilde{r}_A^3H\left(\rho_{eff}
+p_{eff}\right)dt-\frac{\tilde{r}_A}{2\tilde{G}^2}d\tilde{G}\,\,\,.\label{manuel48}
\end{eqnarray}
Defining the work density as \cite{64thermo2}, $W=-(1/2)T^{\alpha\beta(eff)}h_{\alpha\beta}$, one acquires 
\begin{eqnarray}
W=\frac{1}{2}\left(\rho_{eff}-p_{eff}\right)\,\,\,\label{manuel49}.
\end{eqnarray}
Now, using $dV=4\pi \tilde{r}^2_Ad\tilde{r}_A$, one can perform the product $WdV$ as
\begin{eqnarray}
WdV=2\pi\tilde{r}^2_A\left(\rho_{eff}-p_{eff}\right)d\tilde{r}_A\,\,\,. \label{manuel50}
\end{eqnarray}
By summing (\ref{manuel46}) with (\ref{manuel48}), one easily obtains (\ref{manuel50}) plus an additive term, that is $T_HdS = -dE + WdV-T_HdS_{ad}$, which can be written as
\begin{eqnarray}
T_HdS + T_HdS_{ad}= -dE + WdV\,\,\,,\label{manuel51}\\ T_HdS_{ad}=\frac{\tilde{r}_A\left(5-3\omega_{eff}\right)}{8\tilde{G}^2}d\tilde{G}\,\,.\nonumber
\end{eqnarray}
This is the non-equilibrium description of the thermodynamics. This additive term may be interpreted as an entropy production term in the non-equilibrium $R+2g(T)$ gravity. Near the LR, this additive term should be view as the entropy linked with the dissolution of bound structures in LR. This result is quite similar to that obtained in \cite{thermo4,thermo1}. Observe that if $g(T)=0$, the gravitational action is reduced to the Einstein-Hilbert's one and the equilibrium relation is achieved. Specially in this paper, we propose to analyse the behaviour of the entropy in LR model. One can remark from (\ref{manuel43}) that
\begin{eqnarray}
\dot{S}&=&8\pi^2\tilde{r}^3_A\left(\rho_{eff}+p_{eff}\right)-\frac{\pi\tilde{r}^{2}_{A}\dot{\tilde{G}}}{\tilde{G}^2}\nonumber\\
&=&  -\frac{\pi\tilde{r}^2_A}{\tilde{G}}\left(2\tilde{r}_A\dot{H}+\frac{\dot{\tilde{G}}}{\tilde{G}}\right)\, \,\,\,. \label{manuel53}
\end{eqnarray}
By using Eq.~(\ref{manuel26}), Eq.~(\ref{manuel23}), Eq.~(\ref{manuel24}) and the relation $8\pi \tilde{G}=\tilde{\kappa}=\kappa+2g_T$, one gets
\begin{eqnarray}
\tilde{G}=-\frac{\alpha\beta^2}{4\pi(1+\omega)\rho_0}\exp{\left[e^{\beta t}+3\alpha(1+\omega+q)\left(e^{\beta t}-1\right) \right]}\,\,\,, \label{running}
\end{eqnarray}
which means that the running gravitational parameter $\tilde{G}$ is negative in this LR model. Let us show clearly the main reason for which  $\tilde{G}$ is negative. Observe that by using Eq.~(\ref{manuel26}), one gets
\begin{eqnarray}
\kappa+2g_T=-\frac{2\dot{H}}{\rho(1+\omega)}\,\,\,.
\end{eqnarray}
In GR the gravitational action term is just the Einstein-Hilbert term $R$, if the cosmological is note considered. In such a situation, in order to obtain an accelerating expanded universe, one needs to consider an exotic component (the dark energy), with negative pressure such that the parameter of equation of state is negative ($\omega_{DE}<0$). However, when the gravitational action is modified, one acquires the dark energy effect from the contribution of the modified gravity terms, in our case, $2g(T)$. Since this contribution has to play the role of dark energy, just the parameter $\omega_{df}$ associated to the co-called dark fluid may be negative ($\omega_{df}<-1$). One has to consider that the weak energy condition is always satisfied for the ordinary matter, that is $\rho+p>$, or $1+\omega>0$. Moreover, the LR model is guaranteed for $\dot{H}=\ddot{h}>0$. By using these evidences, it follows that the quantity $\kappa+2g_T$ is negative, which means that $\tilde{G}<0$.\par
Moreover, by using (\ref{running}), one acquires
\begin{eqnarray}
\frac{\dot{\tilde{G}}}{\tilde{G}}=\left[\beta+3\alpha(1+\omega+q)\right]e^{\beta t} >0\,\,\,.
\end{eqnarray}
Note that $\dot{H}=\ddot{h}$, and the condition for the occurrence of the LR is $\ddot{h}>0$. Thus, from (\ref{manuel53}), one sees that in the LR universe and the framework of $R+2g(T)$ gravity, the horizon entropy $S$ always increases $\dot{S}>0$.\par In GR where $g(T)=0$, one has $\tilde{G}=G=const$  and the expression (\ref{manuel53}) reduces to 
\begin{eqnarray}
\dot{S}_{GR}= \frac{-2\pi\dot{H}}{GH^3}\,\,\,.\label{entropyGR}
\end{eqnarray}
In this case, for phantom universe ($\dot{H}<0$), the entropy always decreases. This also corresponds to the situation where the null energy condition is violated, i.e. $\rho_{DE}+p_{DE} \leq 0$. This is also occurs in some type of modified gravity \cite{bamba}. In modified gravity the sign of the first derivative of the horizon entropy depends on the form of $f$, i.e. the form of $f(R,T)$, or $f(R)$ or $f(\mathcal{G})$, depending of what type of modified gravity is used ($\mathcal{G}$ is the Gauss-Bonnet invariant term)  \cite{mohseni,12mohseni}. \par
One can now express the horizon entropy corresponding the $R+2g(T)$ model compatible with the LR, as
\begin{eqnarray}
S=\frac{\mathcal{A}}{4\tilde{G}}&=&\frac{8\pi^2}{\tilde{\kappa}H^2}\nonumber\\
&=&-\frac{8\pi^2\rho_0(1+\omega)}{\alpha^3\beta^4a_0^{3(1+\omega+q)}}\exp{\left[-3\beta t-3\alpha(1+\omega+q)\left(e^{\beta t}-1 \right) \right]}\,\,\,,\label{manuel54}\\
&=&\frac{8\pi^2(1+\omega)T}{(3\omega-1)\left[\alpha\beta+k(T)\right]\left[\alpha\beta^2+k(T)\right]}\,\,\,,\quad k(T)=\ln{\left[\left( \frac{T}{T_0} \right)^{-\frac{\beta^2}{3(1+\omega+q)}} \right]}\,\,.
\end{eqnarray}
It is very important to note here that in the framework of GR, the LR cosmology is such that the horizon entropy and its first derivative read
\begin{eqnarray}
S=\frac{\pi}{GH^2}=\frac{\pi}{G\alpha^2\beta^2}e^{-2\beta t}\,\,,\quad \dot{S}=-\frac{2\pi}{G\beta\alpha^2}e^{-2\beta t}<0\,\,\,.
\end{eqnarray}
Once again, we see from this that the behaviour of the entropy quite depend on the form of the gravitational action. Hence, we see that, still guaranteeing the accelerated expansion of the universe, the behaviour of the entropy may change in a LR cosmology, depending of the form of the gravitational action in consideration.  However, in any case, as the time goes to infinity, the entropy vanishes.\par
Without any loss of generality, the assumption (\ref{manuel33}) may be assumed. One can write the horizon entropy around the LR as 
\begin{eqnarray}
S_{LR}&=&S_0\exp{\left[ -3\beta(t_{LR}-t_0)-3\alpha(1+\omega+q)\left(e^{\beta(t_{LR}-t_0)}-1\right)\right]}\,\,\,,\\
S_0&=&-\frac{8\pi^2\rho_0(1+\omega)}{\alpha^3\beta^4a_0^{3(1+\omega+q)}} \,\,\,.
\end{eqnarray}
By using (\ref{manuel33}), one can rewrite the entropy at the LR as 
\begin{eqnarray}
S=S_0\exp{\left\{-3\ln{\left(\frac{H_{LR}}{H_0}\right)}-3\alpha(1+\omega+q)\left[\ln{\left(\frac{H_{LR}}{H_0}\right)}-1\right]\right\} }\,\,\,.\label{entropyLR}
\end{eqnarray}
By considering that the LR must occur very far from today, one can use $H_{LR}>>H_0$ and then, (\ref{entropyLR}) becomes
\begin{eqnarray}
S_{LR}=S_0\left(\frac{H_{LR}}{H_0}\right)^{-3[1+\alpha(1+\omega+q)]}=S_0 e^{3\beta\left[1+\alpha\left(1+\omega+q\right)\right]}\,\,\,.
\end{eqnarray}
Then, since we know the values of the parameters $\alpha$ and $\beta$, the horizon entropy at LR may be calculated. Particularly, if we consider that the ordinary content of the universe is dust, i.e. $\omega=0$, the horizon entropy  at LR reads
\begin{eqnarray}
S_{LR(dust)}= -\frac{3\pi}{\beta GH_0}e^{3(\beta+H_0)}\,\,\,.
\end{eqnarray}
\subsection{Second law of thermodynamics}
In this subsection we consider the Gibbs  equation in terms of all matter and energy fluid  to examine the second law of thermodynamics, that is 
\begin{eqnarray}
T_HdS_{in}=d(\rho_{eff}V)+p_{eff}dV=Vd\rho_{eff}+\left(\rho_{eff}+p_{eff}\right)dV\,\,\,,\label{manuel56}
\end{eqnarray}
where $S_{in}$ denotes the entropy of all the matter and energy inside the horizon. By using the equation of continuity, (\ref{manuel56}) can be written as
\begin{eqnarray}
T_HdS_{in}= -4\pi\tilde{r}^{3}_{A}H\left(\rho_{eff}+p_{eff}\right)+4\pi\tilde{r}^2_{A}\left(\rho_{eff}
+p_{eff}\right)d\tilde{r}_A-\frac{\tilde{r}_A}{2\tilde{G}^2}d\tilde{G}\,\,\,.\label{manuel56in}
\end{eqnarray}
By summing (\ref{manuel56in}) with (\ref{manuel46}), one gets
\begin{eqnarray}
T_H\left( dS+dS_{in}\right)=2\pi\tilde{r}_A\left(\rho_{eff}+p_{eff}\right)d\tilde{r}_A
+\frac{\tilde{r}_A\dot{\tilde{G}}}{4\tilde{G}^2}d\tilde{r}_A-\frac{\tilde{r}_A}{\tilde{G}^2}d\tilde{G}\,\,\,,
\end{eqnarray}
from which, by dividing by $dt$, and using the relation $d\tilde{r}_A/dt=\tilde{\kappa}\left(\rho_{eff}+p_{eff}\right)\tilde{r}^2_A/2$, one gets obtains
\begin{eqnarray}
T_H\left(\dot{S}+\dot{S}_{in}\right)= \frac{\dot{H}^2}{2\tilde{G}H^3}-\frac{\dot{\tilde{G}}}{H\tilde{G}^2}\left( \frac{8\pi \tilde{G}\dot{H}}{H^2}+1\right)\,\,\,.
\end{eqnarray}
The second law is described by
\begin{eqnarray}
\frac{dS}{dt}+\frac{dS_{in}}{dt}\geq 0\,\,\,.\label{manuel57}
\end{eqnarray}
Since the temperature $T_H$ is always positive, one gets the following  condition for the second law of thermodynamics
\begin{eqnarray}
-\frac{\tilde{G}}{\dot{\tilde{G}}}<\frac{2}{\dot{H}}\left(\frac{8}{H}+\frac{H}{\dot{H}}\right)\,\,\,.\label{condition}
\end{eqnarray} 
As we have previously shown,  the quantity $\tilde{G}/\dot{\tilde{G}}$ is always positive in this LR model, and both $H$ and $\dot{H}$ are also positive. Then, (\ref{condition}) always hold for the model $R+2g(T)$, meaning that the second is always satisfied near the LR. By choosing the LR time $t_{LR}\thickapprox H_{LR}^{-1}$ and the present time as $t_0\thickapprox H_{0}^{-1}$, it is easy to observe that $-0.21<e^{-\beta(t_{LR}- t_0)}/\alpha<0.165$, in perfect agreement with the WMAP data \cite{2leonardo}, that is $-1.11<\omega_{DE}<-0.86$. This proves the consistency of the result.

\section{Stability analysis of $R+2g(T)$ LR model}
Let us consider the action {\ref{manuel1}}, with the assumption $f(R,T)=R+2g(T)$, and write the equations (\ref{manuel40}) and (\ref{manuel41}) into the following forms
\begin{eqnarray}
3H^2&=&\rho+\rho_d\,\,\,,\quad\quad\rho_d=\left[2\left(\rho+p\right)g_T+g\right]\,\,\, ,\label{manuel68}\\
\dot{H}&=&-\frac{1}{2}\left(\rho+p+\rho_d+p_d\right)\,\,,\label{manuel69}\\
&=&-\frac{1}{2}\left(1+2g_T\right)\left(\rho+p\right)\,\,\,,\label{manuel70}
\end{eqnarray}
where we set $\kappa=1$.  Let us now define the following dimensionless density parameters
\begin{eqnarray}
x\equiv \frac{\rho_d}{3H^2}\,\,,\quad\quad y\equiv\frac{\rho}{3H^2}\,\,.
\end{eqnarray}
By using the so-called e-folding parameter $N=\ln{a}$, the continuity equations (\ref{deborah24})-(\ref{deborah25}) can be cast into the following forms
\begin{eqnarray}
\frac{dx}{dN}&=&3x(x-1)(1+\omega_d)+3xy(1+\omega)+3qy\,\,\,,\label{manuel74}\\
\frac{dy}{dN}&=&3xy(1+\omega_d)+3y[(1+\omega)(y-1)-q]\,\,\,\label{manuel75}.
\end{eqnarray}
Setting the right hand sides of Eqs. (\ref{manuel74})-(\ref{manuel75}) to zero, one obtains the following critical points: $(i)$ $(x_c,y_c)=(1,0)$ and $(ii)$ $(x_c,y_c)=\left(q/(\omega_d-\omega),(\omega_d-\omega-q)/(\omega_d-\omega)\right)$. Note that the point $(0,0)$ is discarded since, from Eq. (\ref{manuel68}), one must always get $x+y=1$. The critical point $(i)$ corresponds to a universe essentially filled by the dark energy, while the $(ii)$ correspond to a universe  filled by both the ordinary matter and the dark energy.\par
Let us now investigate the stability of the system by considering small perturbations $\delta x$ and $\delta y$ around the  fixed points $(x_c,y_c)$, i.e., $x=x_c+\delta x$ and $y=y_c+\delta y$. Thus, from Eqs. (\ref{manuel74})-(\ref{manuel75}), we obtain the following linearized equations
\begin{eqnarray}
\frac{d(\delta x)}{dN}&=&3\left[(2x_c-1)(1+\omega_d)+y_c(1+\omega)\right]\delta x+3[x_c(1+\omega)+q]\delta y\,\,,\label{manuel76}\\
\frac{d(\delta y)}{dN}&=&3y_c(1+\omega_d)\delta x+3\left[(2y_c-1)(1+\omega)+x_c(1+\omega_d)-q\right]\delta y\,\,\,.\label{manuel77}
\end{eqnarray} 
We can now study the stability of the critical points against perturbations by evaluating the eigenvalues of the matrix corresponding to the system (\ref{manuel76})-(\ref{manuel77}). For the critical point $(i)$, the corresponding eigenvalues are $\lambda_1=3(1+\omega_d)$ and $\lambda_2=3(\omega_d-\omega-q)$, while for the critical point $(ii)$, the eigenvalues are $\lambda_3=3(1+q+\omega)$ and $\lambda_4=3(q+\omega-\omega_d)$.  Let us first analyse the different attribute of the critical point and later, check what of them represent the stable point in LR cosmology in the framework of  $R+2f(T)$ gravity. Observe that, if $-1<\omega_d<\omega+q$, or $\omega+q<\omega_d<-1$, one gets $\lambda_1\lambda_2<0$, and the critical point $(i)$ is a saddle point. For $\omega_d<-1$ and $\omega>\omega_d-q$,   $\lambda_1<0$ and $\lambda_2<0$, meaning that the solution is stable and the critical point is an attractor. For $\omega_d>-1$ and $\omega<\omega_d-q$, both $\lambda_1$ and $\lambda_2$ are positive meaning that the solution is unstable. For the second critical point, one can observe that for $\omega_d<\omega+q<-1$ or $-1<\omega+q<\omega_d$, one has $\lambda_3\lambda_4<0$, meaning that the critical point $(ii)$ is a saddle point. For $\omega>-1-q$ and $\omega-\omega_d>-q$, both $\lambda_3$ and $\lambda_4$ are positive and the corresponding solution is unstable. However, for $\omega<-1-q$ and $\omega_d>\omega+q$, both $\lambda_3$ and $\lambda_4$ are negative, leading to a stable solution and the critical point is an attractor. \par
Here we are interested to the LR model, corresponding to the case where the dark energy dominates because it provides a significant cosmological feature. Let us now analyse the conditions to be satisfied by the input parameters in (\ref{manuel28}) in order to obtain stable solutions. Note that the stability of the LR solution corresponds to $\omega<-1-q$ and $\omega_d>\omega+q$. Since the $\omega$ must always be positive, the realistic  condition for guaranteeing the stability is $\omega+q<\omega_d<-1$, and this is always satisfied for any $q$ such that $q<-1-\omega$. It can be concluded that the stability is realized any $q<-1-\omega$ such that $\omega_d<-1$, and this is always possible since $q$ is an arbitrary non null constant. \par
Remark that from (\ref{manuel68}) $\omega_d$ can be written as follows
\begin{eqnarray}
\omega_d&=&\frac{p_d}{\rho_d}\nonumber\\
&=& \frac{-3H^2-2\dot{H}-p}{3H^2-\rho}\nonumber\\
&=&-1-\frac{2\alpha\beta^2e^{\beta t}+(\omega+1)\rho}{3\alpha^2\beta^2e^{2\beta t}-\rho}\,\,\,,\label{manuel78}
\end{eqnarray}
where we used (\ref{manuel23}). We recall that in LR cosmology the parameter of equation of state $\omega_d$ related to the dark energy is always less than $-1$, because the LR universe is bounded by the $\Lambda$CDM and the BR.  Since $\omega>0$, the numerator of the fraction in (\ref{manuel78}) is always positive. Note also that the denominator $3\alpha^2\beta^2e^{2\beta t}-\rho$ of the same fraction is the energy density of the dark energy of the universe, which is always positive. The LR  (\ref{manuel29}) model obtained in this work is stable and provides a phantom universe without the BR.

\section{Conclusion}
In this paper we reconstructed the cosmological model of effective phantom type and which does not lead to a singularity at finite future time (Little Rip model) in the framework of a type of $f(R, T)$ gravity, where $R$ is the curvature and $T$ the trace of the energy momentum tensor.  Physically, in the LR,  the scale factor and the energy density are never infinite at finite time. The model is reconstructed  by solving an inhomogeneous differential equation of first order from the combination of the generalized Friedmann equations in the special case, $R+2g(T)$, where $g(T)$ is only function of $T$. The integration constant $C_1$ of the general solution is determined by imposing the same initial condition in GR to this modified gravity. Since the initial value of the trace $T_0$ can be directly related to the current Hubble parameter $H_0=2.1\times 0.7\times 10^{-42}$ GeV, the LR models is perfectly reconstructed. By the use of the Supernova Cosmology Project observational data, we show that the parameters $\alpha$ and $\beta$ can be found from the know range of values of the parameter $A$, and then, the input parameters are perfectly in agreement with the observational data.  Furthermore it shown that the model reproduce the $\Lambda$CDM model at this present stage of the universe and confirm that the LR interpolates between the $\Lambda$CDM model and the Big Rip.\par
Moreover, we undertake the thermodynamics in this LR model in the framework of the special $R+2g(T)$ gravity  showing the non-equilibrium description of the thermodynamics. We also observe that  the second law of thermodynamics is always satisfied in this model when the temperature inside the horizon is the same as that on the horizon. An interesting point to be mentioned is that at LR time the current value of $\omega_{DE}$ in this LR model is consistent with the WMAP observational data.\par
In order to test the real validity of the LR  found in this paper, linear perturbations are analysis around the critical point. The result shows that in the current dark energy dominated universe, the model is stable.

\vspace{0.5cm}
{\bf Acknowledgement:} M. J. S. Houndjo thanks  CNPq-FAPES for financial support. M. E. Rodrigues thanks UFES for the hospitality during the elaboration of this work. Authors thank a lot Prof. S. D. Odintsov for the useful comments about stability analysis.

\end{document}